\documentclass[aps,prb,longbibliography,showpacs,reprint,superscriptaddress]{revtex4-1}

\usepackage{graphics}
\usepackage{graphicx}
\usepackage{array}
\usepackage{amsmath}
\usepackage{float}

\newcolumntype{C}[1]{>{\centering\arraybackslash\hspace{0pt}}p{#1}}

\begin{document}

\title{Josephson Metamaterial with a widely tunable positive/negative Kerr constant}

\author{ Wenyuan Zhang}
\affiliation{Department of Physics and Astronomy, Rutgers University, Piscataway, New Jersey 08854}

\author{W. Huang}
\affiliation{Engineering Department, University of Massachusetts Boston, Boston, Massachusetts 02125}

\author{M. E. Gershenson}
\affiliation{Department of Physics and Astronomy, Rutgers University, Piscataway, New Jersey 08854}

\author{M. T. Bell}
\affiliation{Engineering Department, University of Massachusetts Boston, Boston, Massachusetts 02125}

\date{\today}
 
\begin{abstract}
We report on the microwave characterization of a novel one-dimensional Josephson metamaterial composed of a chain of asymmetric superconducting quantum interference devices (SQUIDs) with nearest-neighbor coupling through common Josephson junctions. This metamaterial demonstrates a strong Kerr nonlinearity, with a Kerr constant tunable over a wide range, from positive to negative values, by a magnetic flux threading the SQUIDs.  The experimental results are in good agreement with the theory of nonlinear effects in Josephson chains. The metamaterial is very promising as an active medium for Josephson traveling-wave parametric amplifiers; its use facilitates phase matching in a four-wave mixing process for efficient parametric gain.
\end{abstract}


\maketitle

\section{Introduction}
In conventional optics, a material whose refractive index $n$ is affected by the intensity of an electric field $n \propto |E|^2$ is known as a Kerr medium \cite{Boyd2008}. Analogous to nonlinear optics, microwave superconducting circuits exhibit the Kerr effect due to a nonlinear response of their kinetic/Josephson inductance that determines the circuit impedance. In superconducting circuits based on Josephson junctions the Kerr effect originates from the $\phi^2$ term in the expansion of the Josephson inductance $L(\phi) = \Phi_0/(2\pi I_c\cos \phi)$, where $\phi$ is the superconducting phase across the junction, $I_c$ is the junction critical current, and $\Phi_0$ is the magnetic flux quantum. The Kerr effect in superconducting circuits has been used to generate squeezed states of light \cite{Yurke60}, traveling-wave parametric amplifiers \cite{Macklin307,White24,Bell4,Zorin6}, and superconducting quantum bits \cite{Vlastakis342}.

In this paper, we demonstrate a novel Josephson metamaterial with a Kerr constant tunable over a wide range that includes both positive and negative values. Such a nonlinear medium can find applications in wave-packet rectification \cite{Li4}, analogues of nematic optical materials \cite{Blinov}, superinductances \cite{Bell109}, and in Josephson traveling-wave parametric amplifiers (JTWPA) \cite{Bell4}, which was the motivation behind the present work. The metamaterial is composed of a one-dimensional chain of asymmetric superconducting quantum interference devices (SQUIDs) with nearest-neighbor coupling through common Josephson junctions Fig. \ref{fig1}(a). The same magnetic flux Φ threads all SQUIDs to allow for tunability of the chromatic and nonlinear dispersion. Tunable superconducting metamaterials \cite{Jung27}  composed of passive \cite{Kurter3}, and electrically active meta-atoms such as SQUIDs \cite{Trepanier95,Zhang5,Trepanier3} or flux qubits \cite{Shapiro2} have been investigated. Below we discuss a novel topology with direct coupling between meta-atoms in a structure with a tunable Kerr constant which can change sign. This design offers significant advantages for several applications, including parametric amplification in a JTWPA. As the magnetic flux on the metamaterial is varied, we observe a monotonic dependence of the chromatic dispersion and a Kerr constant which varies over a wide range from positive to negative. This novel metamaterial compares favorably with the Josephson circuits previously used for parametric amplification \cite{Rehak16} in two important aspects. First, the Kerr effect is much stronger and the magnitude of a Kerr constant can be easily tuned by the magnetic flux $\Phi$ in the SQUID loops.  Second, the sign of the Kerr constant is also flux-dependent, which is an important resource for the development of quantum-limited parametric amplifiers and other superconducting circuits. 

\section{Metamaterial Design}
The design of the proposed metamaterial is shown in Fig. \ref{fig1}(a); it resembles the design of the Josephson superinductor introduced by us in Ref. \citenum{Bell109} . Each unit-cell of length $a$ is composed of two coupled asymmetric SQUIDs with a single smaller Josephson junction with critical current $I_{js0}$ and capacitance $C_{js}$ in one arm and two larger Josephson junctions with critical current $I_{jl0}=rI_{js0}$ and capacitance $C_{jl}=rC_{js}$ in the other arm. Here $r$ is the ratio between the areas of the larger and smaller junctions. The field dependent Josephson inductance of the metamaterial is

\begin{align}\label{eq1}
\begin{split}
L(\phi, \Phi)  = L_0 \Big(\Big[\frac{r}{2}+2\cos\Big(2\pi\frac{\Phi}{\Phi_0}\Big)\Big]-\\
\Big[\frac{r}{16}+\cos\Big(2\pi\frac{\Phi}{\Phi_0}\Big)\Big]\phi^2\Big)^{-1},
\end{split}
\end{align}

\noindent where $L_0=\varphi_0/I_{js0}$, $\varphi_0=\Phi_0/(2\pi)$, and $\phi$ is the phase difference across a unit-cell. At a critical value $r_0=4$ the first term in $L(\phi)$ vanishes at $\Phi/\Phi_0=0.5$ and the quadratic term dominates \cite{Bell4,Bell109}. Propagation of electromagnetic waves with wavelengths $\lambda\gg a$ in this metamaterial in the absence of dissipation is described by the following nonlinear wave equation for superconducting phases on the nodes between unit-cells $\varphi(z,t)$ \cite{Bell4,Yaakobi87} 

\begin{align}\label{eq2}
\begin{split}
\frac{a^2}{L_0}\Big[\frac{r}{2}+2\cos\Big(2\pi\frac{\Phi}{\Phi_0}\Big)\Big]\frac{\partial^2\varphi}{\partial z^2}+a^2C_{js}\Big(\frac{r}{2}+2\Big)\frac{\partial^4\varphi}{\partial t^2\partial z^2}-\\C_{gnd}\frac{\partial^2\varphi}{\partial t^2}-\gamma\frac{\partial}{\partial z}\Big[\Big(\frac{\partial\varphi}{\partial z}\Big)^3\Big]=0,
\end{split}
\end{align}

\noindent where $\gamma=a^4/(3\varphi_0^2L_0)[r/16+\cos(2\pi\Phi/\Phi_0)]$. $C_{gnd}$ is the distributed capacitance between the metamaterial and the ground plane. The linear (low-power) dispersion relation and solution to Eq. \eqref{eq2} is

\begin{align}\label{eq3}
\begin{split}
k=\frac{\omega\sqrt{L_0C_{gnd}}}{a\sqrt{\Big[\frac{r}{2}+2\cos\Big(2\pi\frac{\Phi}{\Phi_0}\Big)\Big]-\omega^2L_0C_{js}\Big(\frac{r}{2}+2\Big)}}
\end{split}
\end{align}

\noindent and $A(z)=A_0e^{-i(k+\alpha)z}$ respectively, where

\begin{align}\label{eq4}
\begin{split}
\alpha = \frac{3\gamma k^5|A_0|^2}{8\omega^2C_{gnd}},
\end{split}
\end{align}

\noindent and $A_0$ is the superconducting phase amplitude, see Ref. \citenum{Bell4}. Electromagnetic waves which propagate in this metamaterial acquire a phase shift $-\alpha z$, where $z$ is the direction of propagation along the metamaterial which depends on the intensity $|A_0|^2$ analogous to light traveling in a Kerr medium \cite{Boyd2008,Agrawal2001}. The Kerr constant $\gamma$ and thus the intensity dependent phase shift can vary over a wide range with magnetic flux tuning, and can even change sign from positive to negative.

\section{Microwave Characterization}
\begin{figure}[h!]
\includegraphics[width=8.5cm]{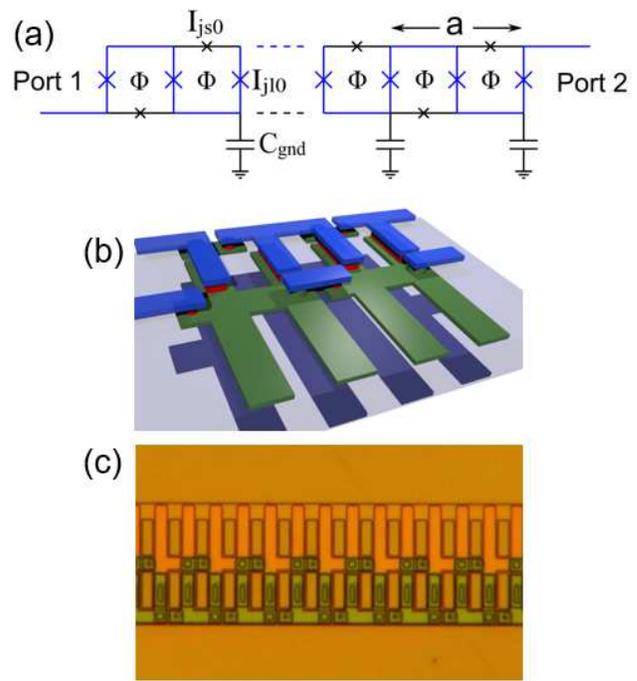}   
\caption{(color online). Josephson metamaterial based on a chain of coupled asymmetric SQUIDs. (a) Circuit schematic of the metamaterial. Each unit-cell of the metamaterial consists of two asymmetric SQUIDs coupled with a shared junction and is of length $a$. Each SQUID in the unit-cell is threaded with a magnetic flux $\Phi$ and has a capacitance to ground $C_{gnd}$. (b) Illustration of the three-metal-layer layout of the device. Metal layer M0 (gray) represents the ground plane, M1 and M2 are the two metal layers which form the electrodes of coupled asymmetric SQUIDs, red and green elements represent Josephson junctions and M1-to-M2 vias respectively. (c) Optical image of the measured Josephson metamaterial. \label{fig1}}
\end{figure}

To demonstrate the tunable properties of the Josephson metamaterial, several devices were fabricated at Hypres Inc. using the standard Nb/AlOx/Nb trilayer process with a nominal critical current density of 30 A/cm$^2$. The devices  are shown schematically in Fig. \ref{fig1}(b). The perforated bottom metal layer M0 (gray) acted as the ground plane; it was separated from the metamaterial structure by 150 nm of SiO$_2$. Metal layers M1 (green) and M2 (blue) form the coupled asymmetric SQUID structure of the metamaterial. The Josephson junctions are shown in red, and the vias between M1 and M2 - in green. Fig. \ref{fig1}(c) shows an optical image of the device. The design parameters of two representative devices are listed in Table 1. The junction critical currents were determined from the Ambegaokar-Baratoff formula \cite{Tinkham} using the normal state resistance $R_{N(s,l)}$ of the on-chip smaller and larger test junctions, respectively, measured at room temperature. The variations in the normal state resistance within the same batch of devices did not exceed 1\%. Each SQUID in the unit-cell has a loop area of $13\times 7 \mu m^2$ and the unit-cell which is composed of two SQUIDs has a length $a = 14 \mu m$. Each device measured contains 125 unit-cells and have a physical length $l=125a$ (1.75 mm). 

\begin{table*}[t]
\centering
\begin{tabular}{ | C{2cm} |  C{1.5cm} | C{1.5cm} | C{1.5cm} | C{1.5cm} | C{1.5cm} | C{1.5cm} | C{1.5cm} | C{1.5cm} | }
\hline
Device & $r$ & $C_{js}$ (fF) & $C_{jl}$ (fF) & $C_{gnd}$ (fF) & $I_{js0}$ ($\mu$A) & $I_{jl0}$ ($\mu$A) & $R_{Ns}$ (k$\Omega$) & $R_{Nl}$ (k$\Omega$)  \\
\hline
1 & 5.9 & 50 & 300 & 75 & 0.25 & 1.5 & 8.44 & 1.42 \\
\hline
2 & 7 & 50 & 350 & 75 & 0.25 & 1.75 & 8.43 &   \\
\hline
  \end{tabular}
\caption{Parameters of two Josephson metamaterial devices.}
\label{table1}
\end{table*}

Investigation of the chromatic and nonlinear dispersion in the Josephson metamaterial was performed in a cryogen-free dilution refrigerator with a base temperature of 20 mK.  The microwave setup with a bandwidth of 1-12 GHz used for characterization is described in Ref. \citenum{Bell86}.  The Josephson chain was included in the microwave transmission line, and transmission measurements were performed with an Anritsu 37369A vector network analyzer. A superconducting solenoid was used to provide a uniform magnetic flux bias to all SQUID loops in the metamaterial.

\begin{figure}[h!]
\includegraphics[width=8.5cm]{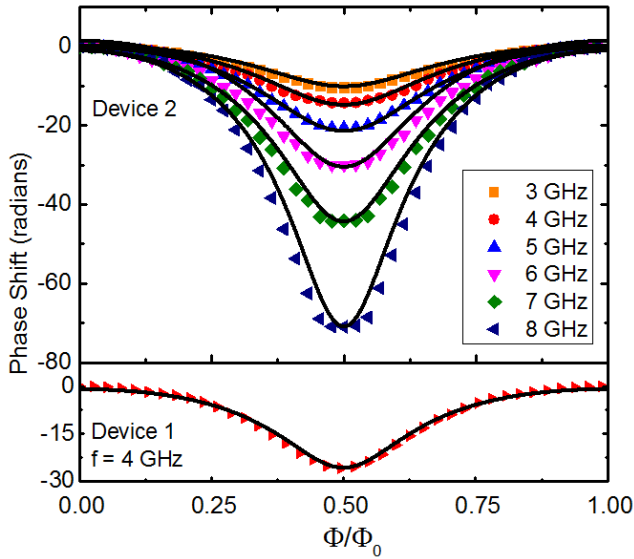}   
\caption{(color online). Low-power transmission measurements of the phase shift across the Josephson metamaterial as a function of the magnetic flux Φ for device 1 (lower panel) and device 2 (upper panel) at different measurement frequencies. Solid lines are fits to Eq. \eqref{eq3}. \label{fig2}}
\end{figure}

\begin{figure}[h!]
\includegraphics[width=8.5cm]{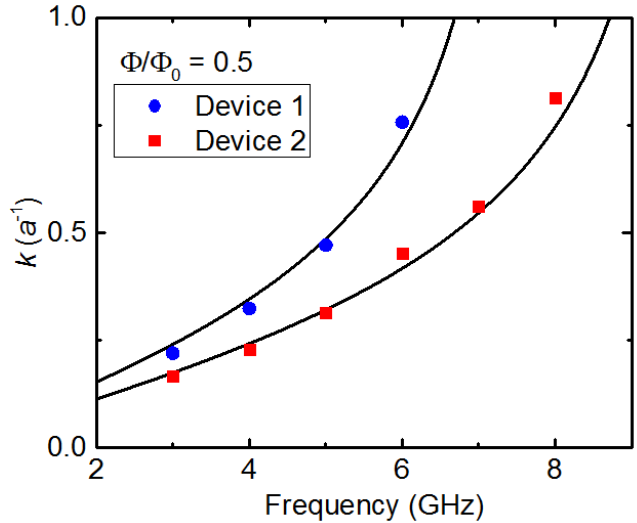}   
\caption{(color online). Wavenumber as a function of frequency for devices 1 (blue circles) and 2 (red squares) extracted from the fitting procedure of the data at $\Phi/\Phi_0=0.5$ in Fig. \ref{fig2}. Solid lines are a plot of Eq. \eqref{eq3} with the design parameters listed in Table \ref{table1}. \label{fig3}}
\end{figure}

The dispersion of the Josephson metamaterial in the linear (low-power) regime was investigated with the transmission measurements of the phase shift as a function of magnetic flux (Fig. \ref{fig2}). The linear transmission measurements were performed at a signal power of ($-130$ dBm) $\textup{--}$ ($-100$ dBm), where $S_{21}$ was independent of the signal intensity. Figure \ref{fig2} illustrates the low-power phase shift across the metamaterial $-lk(\Phi,f)$ measured for devices 1 and 2. Solid lines are fits to Eq. \eqref{eq3} utilizing the design parameters listed in Table \ref{table1} and $I_{js0}$ as the only fitting parameter (for brevity only one measurement frequency $f = 4$ GHz is shown for device 1). The values of $I_{js0}=0.21\pm0.01\mu A$, the same for both devices, were slightly lower than $I_{js0}=0.25 \mu A$ estimated using the Ambegaokar-Baratoff formula (Table \ref{table1}). The chromatic dispersion of the metamaterial at $\Phi/\Phi_0 = 0.5$  is shown for both devices in Fig. \ref{fig3}. Solid lines are the expected $k(f)$ dependence calculated with the initial design parameters in Table \ref{table1}. The effective plasma frequency at $\Phi/\Phi_0 = 0.5$ of the elementary unit-cell is $f_p = [L(\Phi=0.5\Phi_0)(r/2+2)C_{js}]^{-1/2}/2\pi$ which corresponds to 8 GHz and 12 GHz for device 1 and 2, respectively. The phase velocity $\upsilon=a/\sqrt{L(\Phi)C_{gnd}}$ for both devices varied between $3\times10^6$ m/s and $1.5\times10^6$ m/s for $\Phi = 0$ and $\Phi = 0.5\Phi_0$ , respectively. The characteristic impedance $Z=\sqrt{L(\Phi)/C_{gnd}}$ of the metamaterial varied between 60 $\Omega$ and 145 $\Omega$ over the magnetic flux range from 0 to $0.5\Phi_0$.

\begin{figure}[h!]
\includegraphics[width=8.5cm]{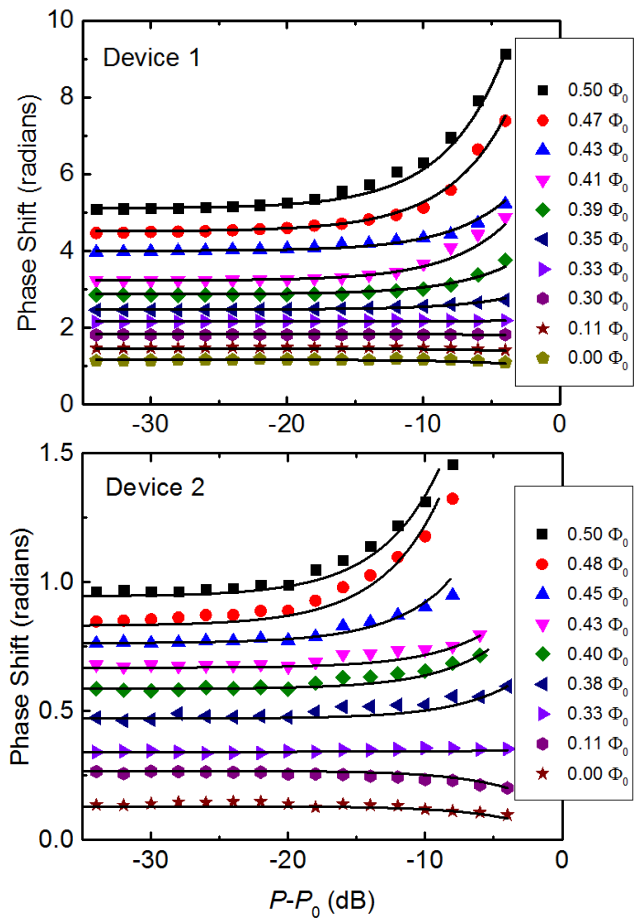}   
\caption{(color online). Measurements of the microwave phase shift as a function of signal power $P$ where $P_0$ = -70 dBm, at different values of the magnetic flux in the metamaterial unit cells for device 1 (upper panel) and device 2 (lower panel). Transmission measurements were performed at a signal frequency of 4 GHz. Solid lines are fits to Eq. \eqref{eq4}. \label{fig4}}
\end{figure}

Figure \ref{fig4} shows the main result of this work: the dependence of the microwave phase shift at a signal frequency of 4 GHz as a function of signal power $P$. An estimated signal power at the mixing chamber of $P_0$ = -70 dBm was attenuated at room temperature with a programmable attenuator (Aeroflex 8311) to vary the signal power $P$ to port 1 of the metamaterial. Microwave transmission measurements were performed over several fixed values of the magnetic flux $0\leq\Phi\leq\Phi_0/2$. For each magnetic flux, the phase across the metamaterial depends on the input power. Near zero field, the phase across the metamaterial decreases with signal power (i.e. a positive Kerr constant), similar to a linear chain of Josephson junctions which would exhibit a non-tunable Kerr constant $\gamma_{JJ}=a^4/(6\varphi_0^2L_0)$ \cite{Yaakobi87}. In contrary to a linear chain of junctions, as the magnetic flux increases, the magnitude and sign of the phase shift changes: $\gamma$ becomes negative at $0.3\leq\Phi/\Phi_0\leq0.5$. The magnitude of the Kerr constant in the metamaterial is similar to $\gamma_{JJ}$ for a linear chain of junctions with critical currents equal to that of the smaller junctions in the metamaterial $\gamma/\gamma_{JJ}=2[r/16+\cos(2\pi\Phi/\Phi_0)]$. However, the metamaterial can be driven with higher microwave power since the majority of the current flows through the “backbone” formed by larger critical current junctions. This feature allows for an increase in the dynamic range of JTWPAs composed of this metamaterial in comparison to linear chains of junctions. According to Eq. \eqref{eq4}  it was expected that the Kerr constant $\gamma$ whould change sign at a magnetic flux of $\Phi=\cos^{-1}(-r/16)/(2\pi)\Phi_0$, which is $\Phi=0.3\Phi_0$ and $\Phi=0.33\Phi_0$ for device 1 and 2 respectively. Indeed, for both devices  the sign change of the Kerr constant was observed in the flux range $0.3-0.33\Phi_0$. In Fig. \ref{fig4} the solid lines are fits to Eq. \eqref{eq4} calculated with the design parameters in Table \ref{table1} and $I_{js0}=0.21 \mu A$. Best fits were obtained with $P_0 = -70$ dBm $\pm 1.5$ dB as a fitting parameter which takes into account the uncertainty of actual signal power level at Port 1 of the metameterial for different tunings of magnetic flux. The nonlinear wave equation (Eq. \eqref{eq2}) which describes the behavior of the Josephson metamaterial is in good agreement with the phase measurement data. 

A nonlinear medium with a tunable Kerr constant, which can change sign, is very promising for the development of Josephson traveling-wave parametric amplifiers \cite{Bell4}. State-of-the-art JTWPAs rely on a four-wave-mixing process which require perfect phase matching between signal, idler and pump waves propagating along a nonlinear transmission line. In recent works \cite{Macklin307,White24,Obrien113,Eom8} the required relations between chromatic dispersion and self- (SPM) and cross-phase (XPM) modulations were realized by tuning the pump frequency near a pole or band-gap introduced into the nonlinear transmission line via sophisticated dispersion engineering techniques. The unique feature of the studied metamaterial enables phase matching due to compensation of  the positive chromatic dispersion between the signal, idler, and pump waves by the negative  SPM and XPM . The theory of operation of such a novel JTWPA was described in Ref. \citenum{Bell4}.

\section{Summary}
In conclusion, we have developed a unique one-dimensional Josephson metamaterial whose Kerr constant is tunable over a wide range, and can change sign from positive to negative. The metamaterial is composed of a chain of coupled asymmetric SQUIDs. The dispersion properties of the metamaterial are varied with an external magnetic flux threading each SQUID loop in the array. The transmission measurements of the phase of microwaves propagating along the metamaterial at low and high signal powers verified predictions of a nonlinear wave equation governing the microwave response  of the medium. Such a metamaterial can be used as the nonlinear medium for parametric amplification and phase-matching in a four-wave-mixing process in Josephson traveling-wave parametric amplifiers, its use eliminates the need for complex dispersion engineering techniques.\\

\begin{acknowledgments}
The work at the University of Massachusetts Boston was supported by NSF awards ECCS-1608448 and DUE-1723511. The work at Rutgers University was supported in part by NSF award DMR-1708954.
\end{acknowledgments}

\bibliography{kerrpaper_ref}

\end{document}